\documentclass[10pt,twocolumn,letterpaper]{article}

\usepackage{cvpr}
\usepackage{times}
\usepackage{epsfig}
\usepackage{graphicx}
\usepackage{amsmath}
\usepackage{amssymb}
\usepackage{mathtools, cuted}
\usepackage{mdframed}
\usepackage[]{subfigure}
\usepackage{multirow}
\usepackage{colortbl}
\usepackage[normalem]{ulem}
\usepackage[table]{xcolor}

\usepackage[export]{adjustbox}
\usepackage{enumitem}

\newif\ifdebugon
\debugontrue

\makeatletter
\newcommand{\Spvek}[2][r]{%
	\gdef\@VORNE{1}
	\left(\hskip-\arraycolsep%
	\begin{array}{#1}\vekSp@lten{#2}\end{array}%
	\hskip-\arraycolsep\right)}

\def\vekSp@lten#1{\xvekSp@lten#1;vekL@stLine;}
\def\vekL@stLine{vekL@stLine}
\def\xvekSp@lten#1;{\def\temp{#1}%
	\ifx\temp\vekL@stLine
	\else
	\ifnum\@VORNE=1\gdef\@VORNE{0}
	\else\@arraycr\fi%
	#1%
	\expandafter\xvekSp@lten
	\fi}
\makeatother

\usepackage[pagebackref=true,breaklinks=true,letterpaper=true,colorlinks,bookmarks=false]{hyperref}

 \cvprfinalcopy 

\def\cvprPaperID{1622} 

\ifcvprfinal\fi
\begin{document}

\title{Computational Imaging for VLBI Image Reconstruction \\ \vspace{0in} }


\author{Katherine L. Bouman$^{1}$
	\and
	Michael D. Johnson$^{2}$
	\and
	Daniel Zoran$^{1}$
	\and
	Vincent L. Fish$^{3}$
	\and
	Sheperd S. Doeleman$^{2,3}$ \hspace{0.4in}
	William T. Freeman$^{1,4}$
	\\ \\
	\small{ \hspace{-0.3in}
		\begin{tabular}{cccc}
			\textsuperscript{1}Massachusetts Institute of Technology, CSAIL  & \textsuperscript{2}Harvard-Smithsonian Center for Astrophysics &
			\textsuperscript{3}MIT Haystack Observatory &
			\textsuperscript{4}Google \\
		\end{tabular}
	}
}

\maketitle

\begin{abstract}
	Very long baseline interferometry (VLBI) is a technique for imaging celestial radio emissions by simultaneously observing a source from telescopes distributed across Earth. 
	The challenges in reconstructing images from fine angular resolution VLBI data are immense. The data is extremely sparse and noisy, thus requiring statistical image models such as those designed in the computer vision community.
	In this paper we present a novel Bayesian approach for VLBI image reconstruction. 
	While other methods often require careful tuning and parameter selection for different types of data, our method (CHIRP) 
	produces good results under different settings such as low SNR or extended emission.
	The success of our method is demonstrated on realistic synthetic experiments as well as publicly available real data. 
	We present this problem in a way that is accessible to members of the community, and provide a dataset website (\url{vlbiimaging.csail.mit.edu}) that facilitates controlled comparisons across algorithms.
\end{abstract}

\vspace{-.2in}
\section{Introduction}
\label{section:introduction}


High resolution celestial imaging is essential for progress in astronomy and physics. 
For example, imaging the plasma surrounding a black hole's event horizon at high resolution could help answer many important questions; 
most notably, it may substantiate the existence of black holes~\cite{blackholesexist} as well as verify and test the effects of general relativity~\cite{nohairtheroem}.  
Recently, there has been an international effort to create an Event Horizon Telescope (EHT) capable of imaging a black hole's event horizon for the first time~\cite{doeleman2012jet, doeleman2008event}. The angular resolution necessary for this observation is at least an order of magnitude smaller than has been previously used to image radio sources~\cite{krichbaum2006sub}.
As measurements from the EHT become available, 
robust algorithms able to reconstruct images in this fine angular resolution regime will be necessary. 

\begin{figure}[t!]
	\centering
	\subfigure[Telescope Locations]{\includegraphics[width=0.45\linewidth]
		{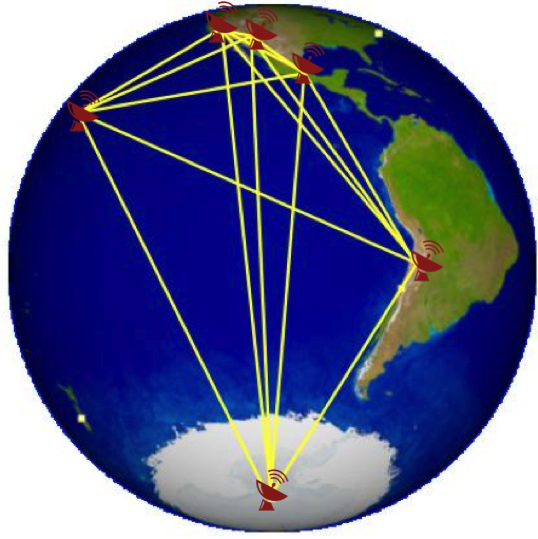}}
	\subfigure[Spatial Frequency Coverage]{\includegraphics[width=0.5\linewidth]
		{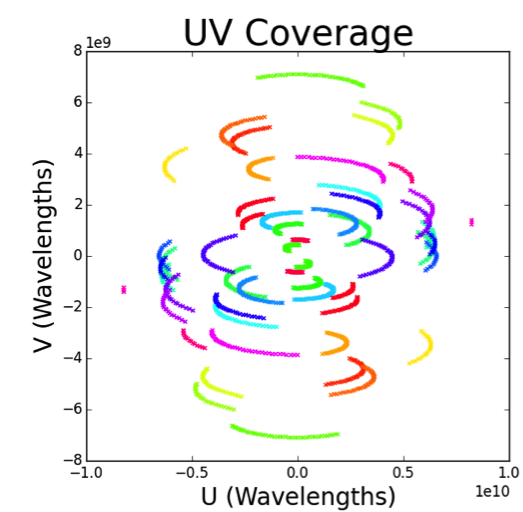}}
	\caption{ \footnotesize{{\bf Frequency Coverage:} (A) A sample of the telescope locations in the EHT. By observing a source over the course of a day, we obtain measurements corresponding to elliptical tracks in the source image's spatial frequency plane (B). These frequencies, $(u,v)$, are the projected baseline lengths orthogonal to a telescope pair's light of sight. Points of the same color correspond to measurements from the same telescope pair.}}
	\label{fig:uvcov}
	\vspace{-0.23in}
\end{figure}

Although billions of dollars are spent on astronomical imaging systems to acquire the best images, current reconstruction techniques suffer from unsophisticated priors and a lack of inverse modeling~\cite{renard2011image}, resulting in sub-optimal images. Image processing, restoration, sophisticated inference algorithms, and the study of non-standard cameras are all active areas of computer vision. The computer vision community's extensive work in these areas are invaluable to the success of these reconstruction methods and can help push the limits of celestial imaging.~\cite{freeman2002example, glasner2009super, levin2011efficient, zoran2011learning}.

Imaging distant celestial sources with high resolving power (i.e. fine angular resolution) requires single-dish telescopes with prohibitively large diameters due to the inverse relationship between angular resolution and telescope diameter~\cite{thompson2008interferometry}.
For example, it is predicted that emission surrounding the black hole at the center of the Milky Way subtends $\approx 2.5\times 10^{-10}$ radians~\cite{fish2014imaging}. Imaging this emission with a $10^{-10}$ radian resolution at a $1.3$ mm wavelength would require a telescope with a $13000$ km diameter. 
Although a single telescope this large is unrealizable, by simultaneously collecting data from an array of telescopes located around the Earth, it is possible to emulate samples from a single telescope with a diameter equal to the maximum distance between telescopes in the array. Using multiple telescopes in this manner is referred to as very long baseline interferometry (VLBI)~\cite{thompson2008interferometry}.  
Refer to Figure~\ref{fig:uvcov}a.

VLBI measurements place a sparse set of constraints on the source image's spatial frequencies. 
The task of reconstructing an image from these constraints is highly ill-posed and relies heavily on priors to guide optimization. 
Current VLBI image reconstruction techniques have been reasonably successful in imaging large celestial sources with coarse angular resolution. 
However, as the demand for higher resolving power increases, particularly for the EHT, traditional reconstruction algorithms are quickly approaching their limits~\cite{rusenimaging, taylor1999synthesis}.

The difficulty of image reconstruction drastically increases as the angular resolution of a VLBI array improves. 
To improve angular resolution (i.e., increase resolving power), one must either increase the maximum distance between two telescopes or decrease the observing wavelength~\cite{thompson2008interferometry}. 
Due to the fixed size of Earth, increasing the maximum telescope baseline results in a smaller set of possible telescope sites to choose from.
Therefore, algorithms must be designed to perform well with increasingly fewer measurements~\cite{rusenimaging}.
Extending VLBI to millimeter and sub-mm wavelengths to increase resolution requires overcoming many challenges, all of which make image reconstruction more difficult.
For instance, at these short wavelengths, 
rapidly varying inhomogeneities in the atmosphere introduce additional measurement errors~\cite{monnier2013radio, taylor1999synthesis}.

In this paper, we leverage ideas from computer vision to confront these challenges. We present a new algorithm, CHIRP (Continuous High-resolution Image Reconstruction using Patch priors), which takes a Bayesian approach and novel formulation to solve the ill-posed inverse problem of image reconstruction from VLBI measurements. Specifically, the contributions of this algorithm are:

\begin{itemize}[leftmargin=*]
	\item \vspace{-.1in} An improved forward model approximation that more accurately models spatial-frequency measurements,
	\item \vspace{-.1in} A simpler problem formulation and optimization strategy to model the effects of atmospheric noise on VLBI data. 
\end{itemize}

\vspace{-.1in}
\noindent{Furthermore, current interferometry testing datasets are small and have noise properties unsuitable for radio wavelengths~\cite{baron20122012, opticalreview, lawson2004interferometry}. }
	\begin{itemize}[leftmargin=*]
	\item  \vspace{-.1in} We introduce a large, realistic VLBI dataset website to the community (\url{vlbiimaging.csail.mit.edu}).
	\end{itemize}
	\vspace{-.1in}
\noindent{This website allows researchers to easily access a large VLBI dataset, and compare their algorithms to other leading methods.
Its automatic evaluation system facilitates unbiased comparisons between algorithms, which are otherwise difficult to make and are lacking in the literature. Furthermore, we hope to make this computational imaging problem accessible to computer vision researchers, cross-fertilizing the astronomy and computer vision communities.
}


\section{A Brief Introduction to VLBI}
\label{sec:vlbi}

We briefly describe VLBI to provide the necessary background for building an accurate likelihood model. Our goal is to provide intuition; for additional details 
we recommend~\cite{thompson2008interferometry}.
As with cameras, a single-dish telescope is diffraction limited. 
However, simultaneously collecting data from an array of telescopes, called an interferometer, allows us to
overcome the single-dish diffraction limit. 

\begin{figure}[tb!]
	\centering
	{\includegraphics[width=\linewidth]
		{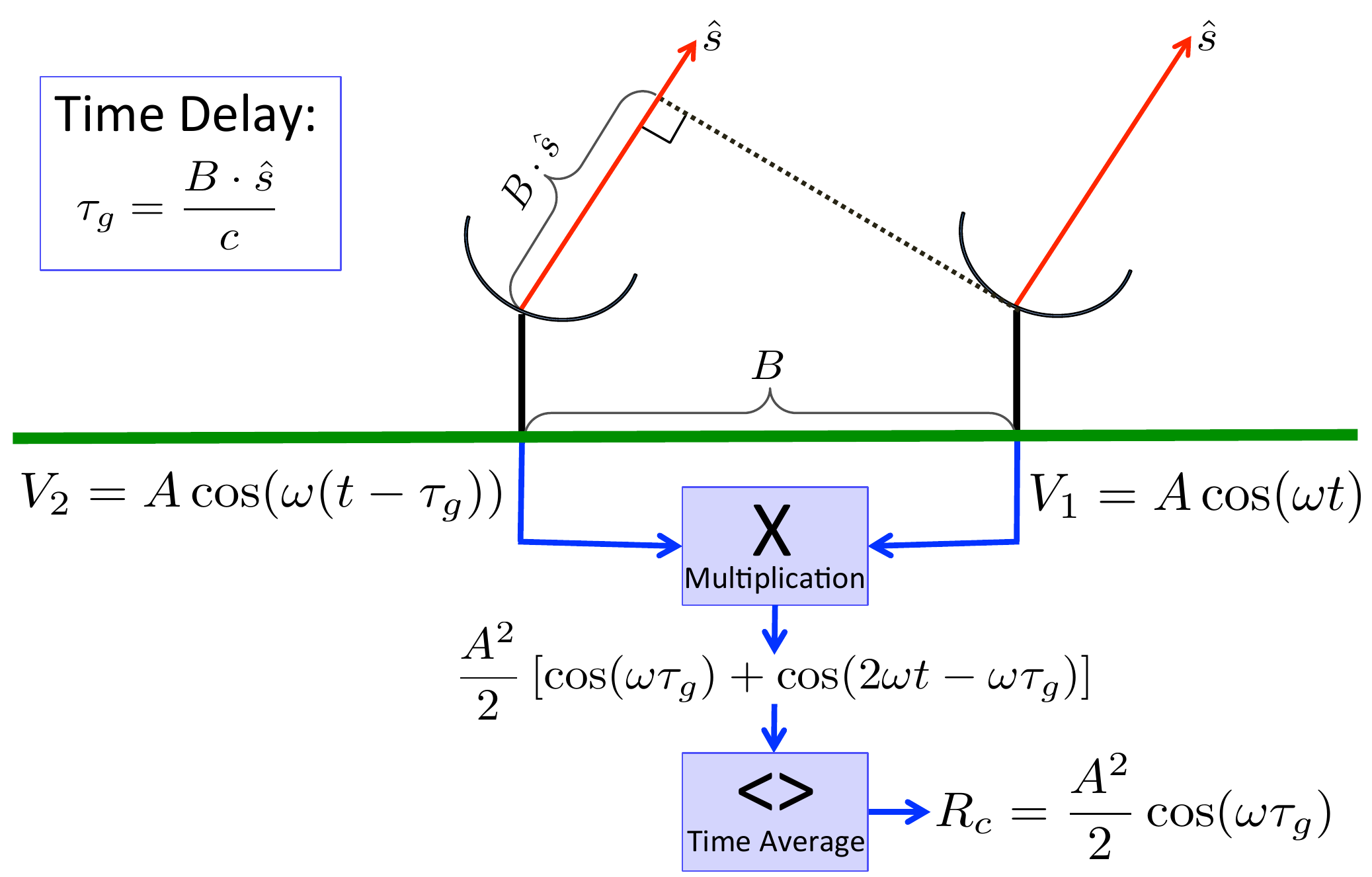}}
	\caption{ \footnotesize{{\bf Simplified Interferometry Diagram:} 
			Light is emitted from a distant source and arrives at the telescopes as a plane wave in the direction $\hat{s}$. An additional distance of $B \cdot \hat{s}$ is necessary for the light to travel to the farther telescope, introducing  a time delay between the received signals that varies depending on the source's location in the sky. The time-averaged correlation of these signals is a sinusoidal function related to the location of the source. This insight is generalized to extended emissions in the van Cittert-Zernike Thm. and used to relate the time-averaged correlation to a Fourier component of the emission image in the direction $\hat{s}$.
			\vspace{-.2in} }}
	\label{fig:introinterferometry}
\end{figure}

Figure~\ref{fig:introinterferometry} provides a simplified explanation of a two telescope interferometer. 
Electromagnetic radiation travels from a point source to the telescopes. However, because the telescopes are separated by a distance $B$, they will not receive the signal concurrently. 
For spatially incoherent extended emissions, the time-averaged correlation of the received signals is equivalent to the projection of this sinusoidal variation on the emission's intensity distribution.


This phenomenon is formally described by the \textit{van Cittert-Zernike Theorem}. The theorem states that, for ideal sensors, the time-averaged correlation of the measured signals from two telescopes, $i$ and $j$, 
for a single wavelength, $\lambda$, can be approximated as:

\vspace{-.2in}
\begin{equation}  \Gamma_{i,j}(u,v) \approx \int_\ell{\int_{m} {e^{-i 2 \pi  (u\ell + vm) }} I_{\lambda}(\ell,m) dl} dm  \label{eq:visibility} \vspace{-.03in} \end{equation}

\noindent{where $ I_{\lambda}(\ell,m)$ is the emission of wavelength $\lambda$ traveling from the direction  $\hat{s} = (\ell, m, \sqrt{1 - \ell^2 - m^2} )$. 
The dimensionless coordinates $(u,v)$ (measured in wavelengths) are the projected baseline, $B$, orthogonal to the line of sight.\footnote{The change in elevation between telescopes can be neglected due to corrections made in pre-processing. Additionally, for small FOVs wide-field effects are negligible.} Notice that Eq.~\ref{eq:visibility} is just the Fourier transform of the source emission image, $I_{\lambda}(\ell,m)$. Thus, $\Gamma_{i,j}(u,v)$
provides a single complex Fourier component of $I_{\lambda}$ at position $(u,v)$ on the 2D spatial frequency plane. {\it We refer to these measurements, $\Gamma_{i,j}$, as visibilities.}
Since the spatial frequency, $(u,v)$, is proportional to the baseline, $B$, increasing the distance between telescopes increases the resolving power of the interferometer, allowing it to distinguish finer details.

\vspace{-0.15in}
\paragraph{Earth Rotation Synthesis} At a single time, for an $N$ telescope array,
we obtain $ \frac{N(N-1)}{2} $ visibility measurements corresponding to each pair of telescopes. 
As the Earth rotates, the direction that the telescopes point towards the source ($\hat{s}$) changes. 
Assuming a static source, this yields measurements of spatial frequency components (visibilities) of the desired image along elliptical paths in the $(u,v)$ frequency plane (see Fig.~\ref{fig:uvcov}b). 

\vspace{-0.15in}
\paragraph{Phase Closure} All equations thus far assumed that light travels from the source to a telescope through a vacuum. However, inhomogeneities in the atmosphere cause the light to travel at different velocities towards each telescope. These delays have a significant effect on the phase of measurements, and renders the phase unusable for image reconstructions at wavelengths less than 3 mm~\cite{monnier2013radio}. 

Although absolute phase measurements cannot be used, a clever observation - termed phase closure - allows us to still recover some information from the phases. 
The atmosphere affects an ideal visibility (spatial frequency measurement) by 
introducing an additional phase term: $\Gamma_{i,j}^{\mbox{\tiny{meas}}} = e^{i(\phi_i - \phi_j)}\Gamma_{i,j}^{\mbox{\tiny{ideal}}}$,
\noindent{where $\phi_i$ and $\phi_j$ are the phase delays introduced in the path to telescopes $i$ and $j$ respectively. By multiplying the visibilities from three different telescopes, we obtain an expression that is invariant to the atmosphere, as the unknown phase offsets cancel, see Eq.~\ref{eq:phaseclosure}~\cite{felli1989very}.  }

\vspace{-.2in}
\begin{align}  
\notag \Gamma^{\mbox{\tiny{meas}}}_{i,j}\Gamma^{\mbox{\tiny{meas}}}_{j,k}\Gamma^{\mbox{\tiny{meas}}}_{k,i} &= e^{i(\phi_i-\phi_j)}\Gamma^{\mbox{\tiny{ideal}}}_{i,j}e^{i(\phi_j-\phi_k)}\Gamma^{\mbox{\tiny{ideal}}}_{j,k}e^{i(\phi_k-\phi_i)}\Gamma^{\mbox{\tiny{ideal}}}_{k,i} \\
&=\Gamma^{\mbox{\tiny{ideal}}}_{i,j}\Gamma^{\mbox{\tiny{ideal}}}_{j,k}\Gamma^{\mbox{\tiny{ideal}}}_{k,i} 
\label{eq:phaseclosure} 
 \end{align}

{\it We refer to this triple product of visibilities as the bispectrum}. The bispectrum is invariant to atmospheric noise; however, in exchange, it reduces the number of constraints that can be used in image reconstruction. Although the number of triple pairs in an $N$ telescope array is ${N\choose 3}$, the number of independent values is only $\frac{(N-1)(N-2)}{2}$.
For small telescope arrays, such as the EHT, this effect is large. 
For instance, in an eight telescope array, using the bispectrum rather than visibilities results in 25\% fewer independent 
constraints 
~\cite{felli1989very}.


 
\vspace{-.05in}
\section{Related Work}
\label{sec:related}
\vspace{-.05in}

We summarize a few significant algorithms from the astronomical interferometry imaging literature.



\vspace{-.15in}
\paragraph{CLEAN}

CLEAN is the de-facto standard method used for VLBI image reconstruction. It assumes that the image is made up of a number of bright point sources. From an initialization image, CLEAN iteratively looks for the brightest point in the image and ``deconvolves" around that location by removing side lobes that occur due to sparse sampling in the $(u,v)$ frequency plane. After many iterations, the final image of point sources is blurred~\cite{hogbom1974aperture}. Since CLEAN assumes a distribution of point sources, it often struggles with reconstructing images of extended emissions~\cite{taylor1999synthesis}.

For mm/sub-mm wavelength VLBI, reconstruction is complicated by corruption of the visibility phases. CLEAN is not inherently capable of handling this problem; however, self-calibration methods have been developed to greedily recover the phases during imaging. Self-calibration requires manual input from a knowledgeable user and often fails when the SNR is too low or the source is complex~\cite{taylor1999synthesis}.




\vspace{-.15in}
\paragraph{Optical Interferometry} 
Interferometry at visible wavelengths faces the same phase-corruption challenges as mm/sub-mm VLBI. 
Although historically the optical and radio interferometry communities have been separate, fundamentally the resulting measurements and imaging process are very similar~\cite{monnier2013radio}. 
We have selected two optical interferometry reconstruction algorithms representative of the field to discuss and compare to in this work~\cite{rusenimaging}. Both algorithms take a regularized maximum likelihood approach and can use the bispectrum, rather than visibilities, for reconstruction~\cite{baron2010novel, buscher1994direct}. Recent methods based on compressed sensing have been proposed, but have yet to demonstrate superior results~\cite{compressedsensing, rusenimaging}. 


 BSMEM (BiSpectrum Maximum Entropy Method) takes a Bayesian approach to image reconstruction~\cite{buscher1994direct}. Gradient descent optimization~\cite{skilling1990quantified} using a maximum entropy prior is used to find an optimal reconstruction of the image. 
 Under a flat image prior BSMEM is often able to achieve impressive super-resolution results on simple celestial images. However, in Section~\ref{section:results} we demonstrate how it often struggles on complex, extended emissions.


SQUEEZE takes a Markov chain Monte Carlo (MCMC) approach to sample images from a posterior distribution~\cite{baron2010novel}. To obtain a sample image, SQUEEZE moves a set of point sources around the field of view (FOV). The final image is then calculated as the average of a number of sample images. 
Contrary to gradient descent methods, SQUEEZE is not limited in its choice of regularizers or constraints~\cite{rusenimaging}. 
However, this freedom comes at the cost of a large number of parameter choices that may be hard for an unknowledgeable user to select and tune. 


\subsection{Spectral Image Reconstruction}

VLBI image reconstruction has similarities with other spectral image reconstruction problems, such as Synthetic Aperture Radar (SAR), Magnetic Resonance Imaging (MRI), and Computed Tomography (CT)~\cite{bracewell2004fourier, lustig2007sparse, 1456966, thibault2007three}. 
However, VLBI faces a number of challenges that are typically not relevant in these other fields. 
For instance, SAR, MRI, and CT are generally not plagued by large corruption of the signal's phase, as is the case due to atmospheric differences in mm/sub-mm VLBI. In SAR the Fourier samples are all coherently related and the absolute phase can generally be recovered, even under atmospheric changes~\cite{533208, 6504845}. However, although fully understanding the connection remains an open problem, incorporating ideas of phase closure, as is done in this work, may open the potential to push SAR techniques~\cite{atmosphereSAR} past their current limits.

\vspace{-.05in}
\section{Method}
\label{section:Method}
\vspace{-.05in}

\begin{figure}[b]
	\vspace{-.0in}
	\centering
	\fcolorbox{red}{red}{\includegraphics[width=0.2\linewidth]{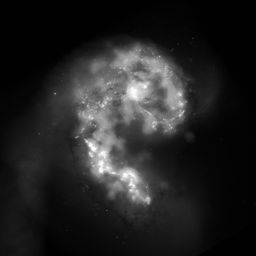}}
	\fcolorbox{green}{green}{\includegraphics[width=0.2\linewidth]{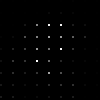}}
	\fcolorbox{cyan}{cyan}{\includegraphics[width=0.2\linewidth]{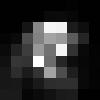}}
	\fcolorbox{blue}{blue}{\includegraphics[width=0.2\linewidth]{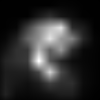}}
	
	\includegraphics[width=0.95\linewidth]{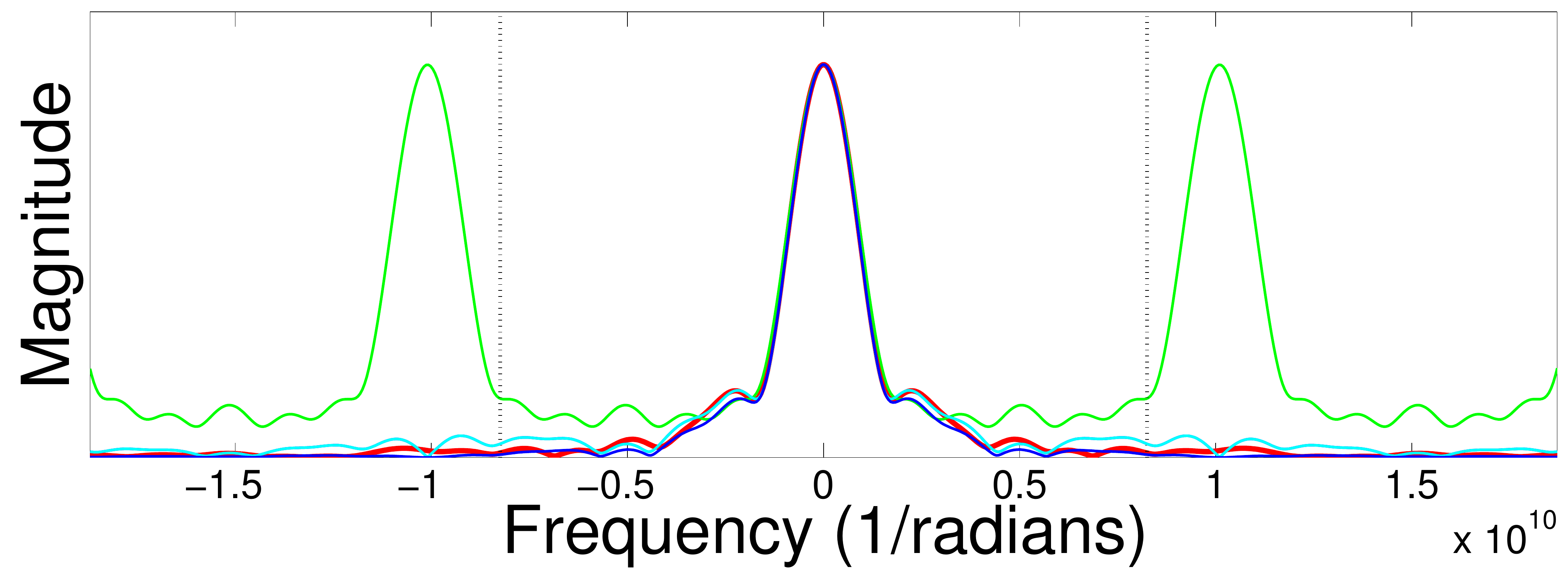}
	\caption{\footnotesize{Accurately modeling the frequencies of an image is crucial for fitting VLBI measurements during image reconstruction. Here we show, that with the same number of parameters, we can much more accurately model the true frequency distribution. A slice of frequencies for the true image is shown in red. Overlayed we show the effect of using the traditional discretized imaging model (green), and our improved model for rectangle (cyan) and triangle (blue) pulses. The dotted lines denote the frequency range sampled in Fig~\ref{fig:uvcov}b. Representing an image in this way reduces modeling errors for higher frequencies during image reconstruction.}}
	\label{fig:pulses}
\end{figure}

\begin{figure*}[!t]
	\vspace{-.2in}
	\normalsize
	\setcounter{equation}{5}
	{\small
		\begin{align}
		\notag \Gamma(u,v) &\approx \int_{- \infty}^{\infty}\int_{- \infty}^{\infty} {e^{-i 2 \pi  (u\ell + vm) }} \sum_{i=0}^{N_\ell-1}\sum_{j=0}^{N_m-1} x[i,j] 
		h \left(\ell- \left( \Delta_{\ell}i + \frac{\Delta_{\ell}}{2}  -\frac{FOV_\ell}{2} \right),m- \left( \Delta_mj + \frac{\Delta_m}{2} -\frac{FOV_m}{2} \right) \right)  d\ell dm  
		\\ &=  \sum_{i=0}^{N_\ell-1}{\sum_{j=0}^{N_m-1}  x[i,j] e^{-i 2 \pi \left(u  \left( \Delta_{\ell}i + \frac{\Delta_{\ell}}{2} + a_\ell \right) + v \left( \Delta_mj + \frac{\Delta_m}{2} + a_m \right)  \right)} H(u, v)} = A {\bf x} = \left( A^{\Re} + i A^{\Im} \right) {\bf x}
		\label{eq:formodel}
		\end{align} 
	}
	\vspace{-.25in}
	\hrulefill
\end{figure*}

Reconstructing an image using bispectrum measurements is an ill-posed problem, and as such there are an infinite number of possible images that explain the data~\cite{rusenimaging}. 
The challenge is to find an explanation that respects our prior assumptions about the ``visual" universe
while still satisfying the observed data. 


\subsection{Continuous Image Representation}

The image that we wish to recover, $I_{\lambda}(\ell,m)$, is defined over the continuous space of angular coordinates $l$ and $m$. 
Many algorithms assume a discretized image of point sources during reconstruction 
~\cite{taylor1999synthesis}. This discretization introduces errors during the reconstruction optimization, especially in fitting the higher frequency visibilities.
Instead, we parameterize a continuous image using a discrete number of terms. 
{\it This parameterization not only allows us to model our emission with a continuous image, but it also reduces modeling errors during optimization. }

Since each measured complex visibility is approximated as the Fourier transform of $I_{\lambda}(\ell,m)$, a convenient parameterization of the image is to represent it as a discrete number of scaled and shifted continuous pulse functions, such as triangular pulses.
For a scene defined in the range $\ell \in [-\frac{F_\ell}{2}, \frac{F_\ell}{2}]$ and $m \in [-\frac{F_m}{2}, \frac{F_m}{2}]$,
we parameterize our space into $N_\ell \times N_m$ scaled pulse functions, $h(l,m)$, centered around 


\begin{align}
 l &= i \Delta_{\ell} + \frac{\Delta_{\ell}}{2}  -\frac{F_\ell}{2} \hspace{0.26in} \mbox{for} \hspace{0.1 in} i = 0,...,N_\ell-1
 \label{eq:discrete1}
\\m &= j\Delta_m + \frac{\Delta_m}{2} -\frac{F_m}{2} \hspace{0.1in} \mbox{for} \hspace{0.1 in} j= 0,...,N_m-1
\label{eq:discrete2}
\end{align}

\vspace{-.09in}
\noindent{for $\Delta_{\ell} = \frac{F_\ell}{N_\ell}$ and $\Delta_m = \frac{F_m}{N_m}$. Using Eq.~\ref{eq:discrete1} and~\ref{eq:discrete2} we can represent a continuous image as a discrete sum of shifted pulse functions scaled by $x[i,j]$. We refer to this image as $\hat{I}_{\lambda}({\bf x})$ for vectorized coefficients ${\bf x}$.

 Due to the shift theorem~\cite{oppenheim1997signals}, plugging this image representation into the van Cittert-Zernike theorem (Eq.~\ref{eq:visibility}) results in a closed-form solution to the visibilities in terms of $H(u,v)$, the Fourier transform of $h(l,m)$, as seen in Eq.~\ref{eq:formodel}. Note that performing a continuous integration has been reduced to a linear matrix operation similar to a Discrete Time Fourier Transform (DTFT).

In Figure~\ref{fig:pulses} we show that this representation allows us to approximates the true frequency components more accurately than a discretized set of point sources, especially for high frequencies.  
Any pulse with a continuous closed-form Fourier transform can be used in this representation (e.g. rectangle, triangle, sinc, Gaussian). The chosen pulse places an implicit prior on the reconstruction. For instance, a sinc pulse with frequency and spacing $\Delta$ can reconstruct any signal with a bandwidth less than $\frac{\Delta}{2}$~\cite{oppenheim1997signals}. In this work we choose to use a triangle pulse with width $(2\Delta_\ell, 2\Delta_m)$ since this is equivalent to linearly interpolating between pulse centers and also simplifies 
non-negativity constraints. 
Although we developed this image model for VLBI image reconstruction, it has potential applicability to a much more general class of imaging problems that rely on frequency constraints.

\vspace{-.05in}
\subsection{Model Energy}
\vspace{-.05in}

We seek a maximum a posteriori (MAP) estimate of the image coefficients, ${\bf x}$, given $M$ complex bispectrum measurements, ${\bf y}$.
Following recent success in image restoration using patch priors~\cite{zoran2011learning, zoran2012natural}, we choose to use an expected patch log likelihood (EPLL) and minimize the energy:

\vspace{-.2in}
\begin{align} 
f_r({\bf x} | {\bf y} )
&=  - D ({\bf y}|{\bf x}) - \mbox{EPLL}_r({\bf x})
\label{eq:bayes}
\end{align}
 \vspace{-.2in}

Eq.~\ref{eq:bayes} appears similar to the familiar form using a Bayesian posterior probability; however, since bispectrum measurements are not independent, the data term $D$ is not a log-likelihood. Additionally, $\mbox{EPLL}_r$ is the expected log likelihood of a randomly chosen patch in the image $\hat{I}_{\lambda}({\bf x})$, not the log-likelihood of the image itself~\cite{zoran2011learning}. 
Details of the data and prior term are discussed below.

\vspace{-.15in}
\subsubsection{Data Term $- D ({\bf y}|{\bf x})$ }

As discussed in Section~\ref{sec:vlbi}, bispectrum measurements are invariant to atmospheric inhomogeneity. 
Therefore, we choose to express an image's ``likelihood" in terms of the bispectrum, rather than visibility, measurements. Let $y_k$ be a noisy, complex bispectrum measurement corresponding to visibilities $k_{1,2}$, $k_{2,3}$, and $k_{3,1}$ for telescopes $1,2$ and $3$. 
Ideal bispectrum values, $\xi$, can be extracted from $\hat{I}_{\lambda}({\bf x})$ using the following polynomial equation:
{\small 
	\begin{align}
	&\xi_k({\bf x}) =  A_{k_{1,2}} {\bf x}  A_{k_{2,3}}{\bf x}  A_{k_{3,1}}{\bf x}  = \xi_k^\Re({\bf x}) + i \xi_k^\Im({\bf x}) 
	\end{align}
}

\vspace{-.2in}
\noindent{where complex, row vector $A_{k_{m,n}}$ extracts the 2D frequency, $(u,v)$, corresponding to the baseline between telescopes $m$ and $n$ from $\hat{I}_{\lambda}({\bf x})$. By assuming Gaussian noise ($\Sigma_k$) on the complex bispectrum measurements, we evaluate $ - D ({\bf y}|{\bf x}) $
	as:}


\vspace{-.15in}

\begin{align} 
\gamma \sum_{i=1}^{M}  & \left[  \frac{\alpha_k}{2}  \Spvek{ \xi_k^\Re({\bf x}) - {\bf y}_k^\Re; \xi_k^\Im({\bf x}) - {\bf y}_k^\Im}^T \Sigma_k^{-1}  \Spvek{ \xi_k^\Re({\bf x}) - {\bf y}_k^\Re;  \xi_k^\Im({\bf x}) - {\bf y}_k^\Im} \right] 
\label{eq:data}
\end{align}


\vspace{-.1in}
\noindent{To account for the independence of each constraint, we set $\alpha_k= \frac{(T_k-1) (T_k-2) }{ 2 {T_k \choose 3} } =  \frac{3}{T_k}$ for $T_k$ telescopes observing at the time corresponding to the $k$-th bispectrum value. If necessary, effects due to the primary beam and interstellar scattering~\cite{fish2014imaging} can be accounted for in Eq.~\ref{eq:data} by weighting each bispectrum term appropriately. Refer to the supp. material~\cite{suppmaterial} for further details.
The polynomial structure of our ``likelihood" formulation is simpler than previously proposed formulations using the bispectrum~\cite{baron2008image, opticalinterferometry}. 

\vspace{-0.15in}
\paragraph{Noise} Although the error due to atmospheric inhomogeneity cancels when using the bispectrum, residual error exists due to thermal noise and gain fluctuations~\cite{thompson2008interferometry}. This introduces Gaussian noise on each complex visibility measurement. Since the bispectrum is the product of three visibilities, its noise distribution is not Gaussian; nonetheless, its noise is dominated by a Gaussian bounding its first-order noise terms. In the supp. material we show evidence that this is a reasonable approximation~\cite{suppmaterial}.



\vspace{-.15in}
\subsubsection{Regularizer $- \mbox{EPLL}_r({\bf x})$ }


We use a Gaussian mixture model (GMM) patch prior to regularize our solution of $\hat{I}_{\lambda}({\bf x})$. Although this is not an explicit prior on $\hat{I}_{\lambda}({\bf x})$, patch priors have been shown to work well 
in the past~\cite{zoran2011learning, zoran2012natural}.  Building on ideas from~\cite{zoran2011learning}, we maximize the EPLL by maximizing the probability of each of the $N$ overlapping pulse patches in $\hat{I}_{\lambda}({\bf x})$:

\vspace{-.15in}
\begin{align}
\mbox{EPLL}_r({\bf x}) = \textstyle \sum_{n=1}^N  \log p(P_n {\bf x}).
\end{align}
\vspace{-.15in}

\noindent{$P_n$ is a matrix that extracts the $n$-th patch from {\bf x} and $p(P_n {\bf x})$ is the probability of that patch learned through the GMM's optimization. We use a patch size of 8x8.}





\subsection{Optimization}

\noindent{To optimize Eq.~\ref{eq:bayes} we use ``Half Quadratic Splitting"~\cite{zoran2011learning}. This method introduces a set of auxiliary patches $\{ z^i \}_1^N$, one for each overlapping patch $P_i{\bf x}$ in the image. We can then solve this problem using an iterative framework:




\vspace{0.05in}
\noindent{{\bf (1) Solve for $ \{ z^n \}$ given ${\bf x}$}: In order to complete this step we set  $ \{ z^n \}$ to the most likely patch under the prior, given the corrupted measurements $P_nX$ and weighting parameter $\beta$ (described further in~\cite{zoran2011learning}).}

\vspace{0.05in}
\noindent{{\bf (2) Solve for ${\bf x}$ given $ \{ z^n \}$}: If we were able to work with visibilities our problem would be quadratic in ${\bf x}$, and we could solve then for ${\bf x}$ in closed-form. However, since we use bispectrum measurements rather than visibilities, our energy is a 6th order polynomial that cannot be solved in closed-form. One possibility is to solve for the gradient and then use gradient descent to solve for a local minimum. However, this method is slow. Alternatively, we perform a 2nd order Taylor expansion around our current solution, ${\bf x}_0$, to obtain an approximation of the local minimum in closed-form. A detailed derivation of the gradient and local closed-form solution can be found in the supp. material~\cite{suppmaterial}. 

As suggested in~\cite{zoran2011learning} we iterate between these two steps for increasing $\beta$ values of $1, 4, 8, 16, 32, 64, 128, 256,$ $512$.  


\vspace{-.15in}
\paragraph{Multi-scale Framework}
Since our convexification of the energy is only approximate, we slowly build up $\hat{I}_{\lambda}({\bf x})$ using a multi scale framework. This framework also helps to avoid local minima in the final solution. We initialize the image ${\bf x}_0$ with small random noise centered around the mean flux density (average image intensity), and iterate between solving for $ \{ z^n \}$ and ${\bf x}$. Then, using our discretized formulation of the image, we increase the number of pulses used to describe the image. This framework allows us to find the best low-resolution solution before optimizing the higher frequency detail in the image. In this paper, we initialize our optimization using a set of $20 \times 20$ pulses and slowly increase to $64 \times 64$ pulses over $10$ scales.

\section{Dataset}
\label{section:dataset}

We introduce a dataset and website (\url{vlbiimaging.csail.mit.edu}) for evaluating the performance of VLBI image reconstruction algorithms. 
By supplying a large set of easy-to-understand training and testing data, we hope to make the problem more accessible to those less familiar with the VLBI field. 
The website contains a:

\begin{itemize}[leftmargin=*]
\item  \vspace{-.1in} Standardized data set of real and synthetic data for training and blind testing of VLBI imaging algorithms
\item  \vspace{-.1in} Automatic quantitative evaluation of algorithm performance on realistic synthetic test data
\item  \vspace{-.1in} Qualitative comparison of algorithm performance 
\item  \vspace{-.1in} Online form to easily simulate realistic data using user-specified image and telescope parameters \vspace{-.05in}
\end{itemize}



Current interferometry dataset challenges are both small
and assume noise characteristics unsuitable for radio wavelengths~\cite{baron20122012, lawson2004interferometry}.
The introduction of this new radio VLBI dataset will help to reveal shortcomings of current methods as well as encourage the development of new algorithms. 


\vspace{-.2in}
\paragraph{Synthetic Measurements}

We provide a standardized format~\cite{ pauls2005data} dataset of over 5000 synthetic VLBI measurements corresponding to
a variety of array configuration, source images, and noise levels. 
Measurements are simulated using the MIT Array Performance Simulator (MAPS) software package~\cite{rusenimaging}.
This software has been developed to accurately model the visibilities and noise expected from a user-specified array and source. {\it Visibilities from the MAPS are not generated in the same manner as our forward model.} 

We generate data using a collection of black hole~\cite{avery}, celestial~\cite{jpl, nrao}, and natural images. 
We have deliberately included a diversity of images in the imaging database, since imaging algorithms for black holes must be sufficiently non-committal that they can identify departures from canonical expectations. 
Natural images test robustness to complex scenes with varied image statistics.

\vspace{-.15in}
\paragraph{Real Measurements} We provide 33 sets of measurements from the VLBA-BU-BLAZAR Program~\cite{jorstad2005polarimetric} in the same standardized format~\cite{ pauls2005data}. 
This program has been collecting data on a number of gamma-ray blazars every month since 2007. 
Measurements are taken using the Very Long Baseline Array (VLBA) at 43 GHz. Although both the angular resolution and wavelength of these measurements are very different from those taken by the EHT (which collects at $\approx 230$ GHz)~\cite{fish2014imaging}, they provide a means to test algorithms on measured, experimental data. 

\vspace{-.15in}
\paragraph{Test Set and Error Metrics}

We introduce a blind test set of
challenging synthetic data. 
Measurements with realistic errors are generated using a variety of target sources and telescope parameters and provided in the OIFITS format~\cite{pauls2005data}.
This test set introduces a means for fair quantitative comparisons between competing algorithms. Researchers are encouraged to run their algorithms on this data and submit results to the website for evaluation.

Traditional point-by-point error metrics, such as MSE and PSNR, are sometimes uninformative in the context of highly degraded VLBI reconstructions. 
Therefore, we supplement the MSE metric with the perceptually motivated structural similarity (SSIM) index~\cite{wang2003ssim}.
Since the absolute position of the emission is lost when using the bispectrum, we first align the reconstruction to the ground truth image using cross-correlation. We then evaluate the MSE and SSIM on the normalized, aligned images. 
	Although we consider MSE and SSIM a good first step towards quantitative analysis, we believe a better metric of evaluation is subject for future research. 



\vspace{-.1in}
\section{Results and Discussion}
\label{section:results}
\vspace{-.05in}

\begin{figure}[b]
	\begin{center}
		\vspace{-.2in}
		\begin{tabular}{  c  c  c  c  c  }

			\multirow{1}{*}[0.4in]{ \rotatebox[origin=t]{90}{{\textsf{Source}} }} &
			\includegraphics[width=.17\linewidth]
			{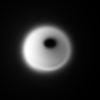} &
			\includegraphics[width=.17\linewidth]
			{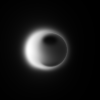} & \includegraphics[width=.17\linewidth]
			{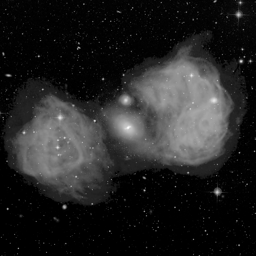} & 
			\includegraphics[width=.17\linewidth]
			{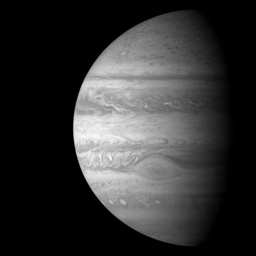}
			\\

			\hline
			&\vspace{-.1in}&&&\\
			\multirow{1}{*}[0.45in]{ \rotatebox[origin=t]{90}{{\textsf{Max Res}} }} &
			\includegraphics[width=.17\linewidth]
			{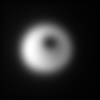} &
			\includegraphics[width=.17\linewidth]
			{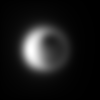} & \includegraphics[width=.17\linewidth]
			{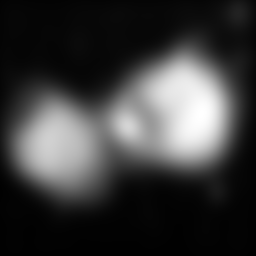} & 
			\includegraphics[width=.17\linewidth]
			{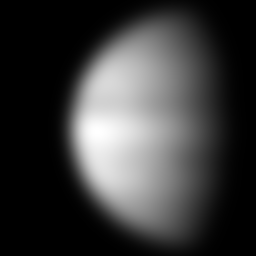}
			\\

		\end{tabular}
		\caption{ \footnotesize{{\bf Intrinsic Maximum Resolution:} The geometry of a telescope array imposes an intrinsic maximum resolution on the image you can reconstruct from the measurements. Recovering spatial frequencies higher than this resolution is equivalent to superresolution. For results presented, the minimum recoverable fringe spacing (corresponding to the maximum frequency) is 24.72 $\mu$-arcseconds. The original `Source' images (183.82 $\mu$-arcsecond FOV) are used to synthetically generate realistic VLBI measurements. We show the effect of filtering out spatial frequencies higher than the minimum fringe spacing for these source images in `Max Res'. }}
		\label{fig:maxres}
		\vspace{-.3in}
	\end{center}
\end{figure}

	\begin{figure*}[ht!]
		\vspace{-0.2in}
		\begin{center}
			\begin{tabular}{  c | c  c |  c  c  c  c | c  }

				& \multicolumn{2}{c}{ \small{\textsf{BLACK HOLE}} }  &   \multicolumn{4}{c}{ \small{\textsf{CELESTIAL}} } &  \small{\textsf{NATURAL}}  \\ \hline

				\vspace{-.1in} \cellcolor[gray]{0.8} & \cellcolor[gray]{0.8}& \cellcolor[gray]{0.8}& \cellcolor[gray]{0.8}& \cellcolor[gray]{0.8}&\cellcolor[gray]{0.8} & \cellcolor[gray]{0.8} & \cellcolor[gray]{0.8} \\
				
				\cellcolor[gray]{0.8} \multirow{1}{*}[.5in]{ \rotatebox[origin=t]{90}{{\textsf{TARGET}}} } &
				\cellcolor[gray]{0.8}\includegraphics[width=.1\linewidth]
				{blackhole40_fliltered}  &
				\cellcolor[gray]{0.8}\includegraphics[width=.1\linewidth]
				{blackhole_filtered.png} &
				\cellcolor[gray]{0.8}\includegraphics[width=.1\linewidth]
				{celestial-03-20-1.png} & \cellcolor[gray]{0.8}\includegraphics[width=.1\linewidth]
				{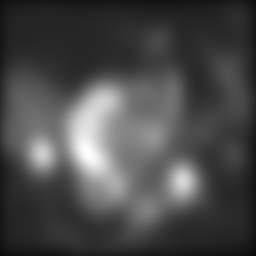}& \cellcolor[gray]{0.8}\cellcolor[gray]{0.8}\includegraphics[width=.1\linewidth]
				{celestial-18-20-1.png}& \cellcolor[gray]{0.8}\cellcolor[gray]{0.8}\includegraphics[width=.1\linewidth]
				{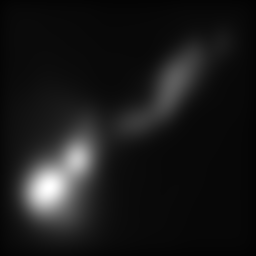}& \cellcolor[gray]{0.8}\cellcolor[gray]{0.8}\includegraphics[width=.1\linewidth]
				{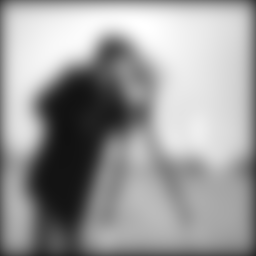} \\
				\hline
				
			\vspace{-.1in}& &&&& & & \\
				
				\multirow{1}{*}[0.5in]{ \rotatebox[origin=t]{90}{{\textsf{CLEAN}}} } &
				\includegraphics[width=.1\linewidth]
				{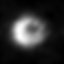}  &
				\includegraphics[width=.1\linewidth]
				{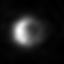} &
				\includegraphics[width=.1\linewidth]
				{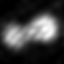} & \includegraphics[width=.1\linewidth]
				{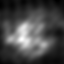}& \includegraphics[width=.1\linewidth]
				{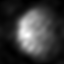}& \includegraphics[width=.1\linewidth]
				{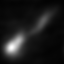}& \includegraphics[width=.1\linewidth]
				{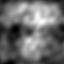} \\
				\hline
				
				\vspace{-.1in}& &&&& & & \\
				\multirow{1}{*}[0.55in]{ \rotatebox[origin=t]{90}{{\textsf{SQUEEZE}}} }&
				\includegraphics[width=.1\linewidth]
				{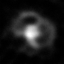}  &
				\includegraphics[width=.1\linewidth]
				{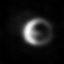} &
				\includegraphics[width=.1\linewidth]
				{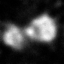} & \includegraphics[width=.1\linewidth]
				{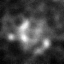}& \includegraphics[width=.1\linewidth]
				{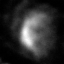}& \includegraphics[width=.1\linewidth]
				{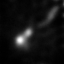}& \includegraphics[width=.1\linewidth]
				{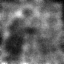} \\
				\hline
				
				\vspace{-.1in}& &&&& & & \\
				\multirow{1}{*}[0.5in]{ \rotatebox[origin=t]{90}{{\textsf{BSMEM}}} } &
				\includegraphics[width=.1\linewidth]
				{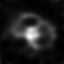}  &
				\includegraphics[width=.1\linewidth]
				{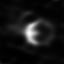} &
				\includegraphics[width=.1\linewidth]
				{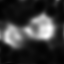} & \includegraphics[width=.1\linewidth]
				{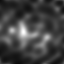}& \includegraphics[width=.1\linewidth]
				{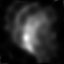}& \includegraphics[width=.1\linewidth]
				{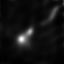}& \includegraphics[width=.1\linewidth]
				{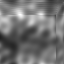} \\
				\hline
				
				\vspace{-.1in}\cellcolor[gray]{0.8} & \cellcolor[gray]{0.8}& \cellcolor[gray]{0.8}& \cellcolor[gray]{0.8}& \cellcolor[gray]{0.8}&\cellcolor[gray]{0.8} & \cellcolor[gray]{0.8} & \cellcolor[gray]{0.8} \\
				
				\cellcolor[gray]{0.8} \multirow{1}{*}[0.5in]{ \rotatebox[origin=t]{90}{{\textsf{CHIRP}}} } &
				\cellcolor[gray]{0.8}\includegraphics[width=.1\linewidth]
				{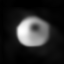}  &
				\cellcolor[gray]{0.8}\includegraphics[width=.1\linewidth]
				{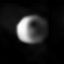} &
				\cellcolor[gray]{0.8}\includegraphics[width=.1\linewidth]
				{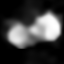} & \cellcolor[gray]{0.8}\includegraphics[width=.1\linewidth]
				{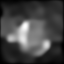}& \cellcolor[gray]{0.8}\includegraphics[width=.1\linewidth]
				{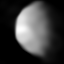}& \cellcolor[gray]{0.8}\includegraphics[width=.1\linewidth]
				{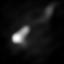}& \cellcolor[gray]{0.8}\includegraphics[width=.1\linewidth]
				{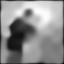} \\
				\hline
\vspace{-.1in}& &&&& & & \\
				& \small{\textsf{A}}  & \small{\textsf{B}}  & \small{\textsf{C}}  & \small{\textsf{D}} & \small{\textsf{E}} & \small{\textsf{F}} &  \small{\textsf{G}}  \\

			\end{tabular}
			\caption{\footnotesize{ {\bf Method Comparison: } Comparison of our algorithm, `CHIRP' to three state-of-the-art methods: `CLEAN', `SQUEEZE', and `BSMEM'. We show the normalized reconstruction of a variety of black hole (a-b), celestial (c-f), and natural (g) source images with a total flux density (sum of pixel intensities) of 1 jansky and a 183.82 $\mu$-arcsecond FOV. Since absolute position is lost when using the bispectrum, shifts in the reconstructed source location are expected. The `TARGET' image shows the ground truth emission filtered to the maximum resolution intrinsic to this telescope array.  \vspace{-.2in}  
				}}
				\label{fig:comp}
			\end{center}
		\end{figure*}

	\begin{figure*}[t]
		\begin{center}
			\vspace{-.0in}
			\begin{tabular}{  c | c | c | c | c  }
				& \footnotesize{\textsf{CLEAN}}  & \footnotesize{\textsf{SQUEEZE}}  & \footnotesize{\textsf{BSMEM}} &  \cellcolor[gray]{0.8}\footnotesize{\textsf{CHIRP}}  \\ \hline
				&\vspace{-.1in}&&&\cellcolor[gray]{0.8}\\
				
				\multirow{1}{*}[0.4in]{ \rotatebox[origin=t]{90}{{\textsf{3.0 Flux}} }} &
				\includegraphics[width=.08\linewidth]
				{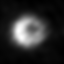} &
				\includegraphics[width=.08\linewidth]
				{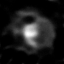} & \includegraphics[width=.08\linewidth]
				{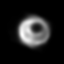} & 
				\cellcolor[gray]{0.8}\includegraphics[width=.08\linewidth]
				{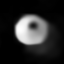}
				\\

				\hline
				&\vspace{-.1in}&&&\cellcolor[gray]{0.8}\\
				\multirow{1}{*}[0.4in]{ \rotatebox[origin=t]{90}{{\textsf{2.0 Flux}} }} &
				\includegraphics[width=.08\linewidth]
				{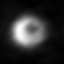} &
				\includegraphics[width=.08\linewidth]
				{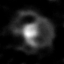} & \includegraphics[width=.08\linewidth]
				{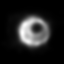} & 
				\cellcolor[gray]{0.8}\includegraphics[width=.08\linewidth]
				{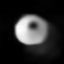}
				\\

				\hline
				&\vspace{-.1in}&&&\cellcolor[gray]{0.8}\\
				\multirow{1}{*}[0.4in]{ \rotatebox[origin=t]{90}{{\textsf{0.5 Flux}} }} &
				\includegraphics[width=.08\linewidth]
				{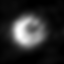} &
				\includegraphics[width=.08\linewidth]
				{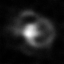} & \includegraphics[width=.08\linewidth]
				{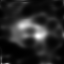} & 
				\cellcolor[gray]{0.8}\includegraphics[width=.08\linewidth]
				{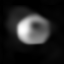}
				\\

			\end{tabular}
			\quad
			\begin{tabular}{  c | c | c | c | c  }
				& \footnotesize{\textsf{CLEAN}}  & \footnotesize{\textsf{SQUEEZE}}  & \footnotesize{\textsf{BSMEM}} &  \cellcolor[gray]{0.8}\footnotesize{\textsf{CHIRP}}  \\ \hline
				&\vspace{-.1in}&&&\cellcolor[gray]{0.8}\\
				
				\multirow{1}{*}[0.4in]{ \rotatebox[origin=t]{90}{{\textsf{3.0 Flux}} }} &
				\includegraphics[width=.08\linewidth]
				{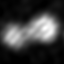} &
				\includegraphics[width=.08\linewidth]
				{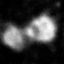} & \includegraphics[width=.08\linewidth]
				{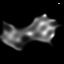} & 
				\cellcolor[gray]{0.8}\includegraphics[width=.08\linewidth]
				{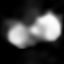}
				\\

				\hline
				&\vspace{-.1in}&&&\cellcolor[gray]{0.8}\\
				\multirow{1}{*}[0.4in]{ \rotatebox[origin=t]{90}{{\textsf{2.0 Flux}} }} &
				\includegraphics[width=.08\linewidth]
				{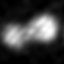} &
				\includegraphics[width=.08\linewidth]
				{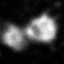} & \includegraphics[width=.08\linewidth]
				{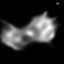} & 
				\cellcolor[gray]{0.8}\includegraphics[width=.08\linewidth]
				{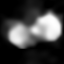}
				\\

				\hline
				&\vspace{-.1in}&&&\cellcolor[gray]{0.8}\\
				\multirow{1}{*}[0.4in]{ \rotatebox[origin=t]{90}{{\textsf{0.5 Flux}} }} &
				\includegraphics[width=.08\linewidth]
				{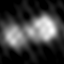} &
				\includegraphics[width=.08\linewidth]
				{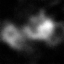} & \includegraphics[width=.08\linewidth]
				{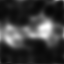} & 
				\cellcolor[gray]{0.8}\includegraphics[width=.08\linewidth]
				{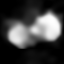}
				\\

			\end{tabular}
			\caption{ \footnotesize{{\bf Noise Sensitivity:} The effect of varying total flux density (in janskys), and thus noise, on each method's recovered reconstructions. Decreasing flux results in higher noise. Notice how our method is fairly robust to the noise, while the results from other methods often vary substantially across the noise levels. The ground truth target images along with the results for a total flux density of 1 jansky can been seen in column A and C of Figure~\ref{fig:comp}. 
					}}
			\label{fig:noise}
			\vspace{-.3in}
		\end{center}
	\end{figure*}

Measurements from the EHT have yet to become available.
Therefore, we demonstrate the success of our algorithm, CHIRP, on a sample of synthetic examples obtained from our online dataset and real VLBI measurements collected by the VLBA-BU-BLAZAR Program.

\vspace{-.15in}
\paragraph{Synthetic Measurements}
For image results presented in the paper we generated synthetic data using realistic parameters for the EHT array pointed towards the black hole in M87.
Corresponding $(u,v)$ frequency coverage is shown in Figure~\ref{fig:uvcov}b.
The geometry of an array imposes an intrinsic maximum resolution on the image you can reconstruct from its measurements. Figure~\ref{fig:maxres} shows the effect of filtering out spatial frequencies higher than the minimum fringe spacing.
These images set expectations on what is possible to reliably reconstruct from the measurements.
Additional results and details about the telescopes, noise properties, and parameter settings can be found in the supp. material~\cite{suppmaterial}. 

\vspace{-.15in}
\paragraph{Method Comparison} We compare results from our algorithm, CHIRP, with the three state-of-the-art algorithms described in Section~\ref{sec:related}: CLEAN, SQUEEZE, and BSMEM. 
Images were obtained by asking authors of the competing algorithms or knowledgeable users for a suggested set of reconstruction parameter (provided in the supp. material~\cite{suppmaterial}). 
The website submission system allows the results from other parameter settings and
algorithms to be compared, both qualitatively and quantitatively.

As with our algorithm, SQUEEZE~\cite{baron2010novel} and BSMEM~\cite{buscher1994direct} use the bispectrum as input. 
CLEAN cannot automatically handle large phase errors, so CLEAN results were obtained using {\it calibrated} (eg. no atmospheric phase error) visibilities in CASA~\cite{jaeger2008common}. In reality, these ideal calibrated visibilities would not be available, and the phase would need to be recovered through highly user-dependent self-calibration methods. However, in the interest of a fair comparison, we show the results of CLEAN in a ``best-case" scenario. 

Figure~\ref{fig:comp} shows a sample of results comparing our reconstructions to those of the current state-of-the-art methods. Our algorithm is able to handle a wide variety of sources, ranging from very simple celestial to complex natural images, without any additional parameter tuning. 
CLEAN produces consistently blurrier results. 
Both SQUEEZE and BSMEM tend towards sparser images. This strategy works well for superresolution.
However, it comes at the cost of often making extended sources overly sparse and introducing spurious detail. 
Although algorithms such as BSMEM and SQUEEZE may perform better on these images with specific hand-tuned parameters, these tests demonstrate that the performance of CHIRP requires less user expertise and provides images that may be less sensitive to user bias.

Figure~\ref{fig:resultnums} shows a quantitative comparison of our method to SQUEEZE and BSMEM for the challenging, blind test set presented in Section~\ref{section:dataset}. 
Since CLEAN cannot automatically handle large phase errors, we were unable to fairly compare its results on this test set. 

\begin{figure}[t!]
		\vspace{-.25in}
	\centering
	\subfigure{\includegraphics[width=0.49\linewidth]
		{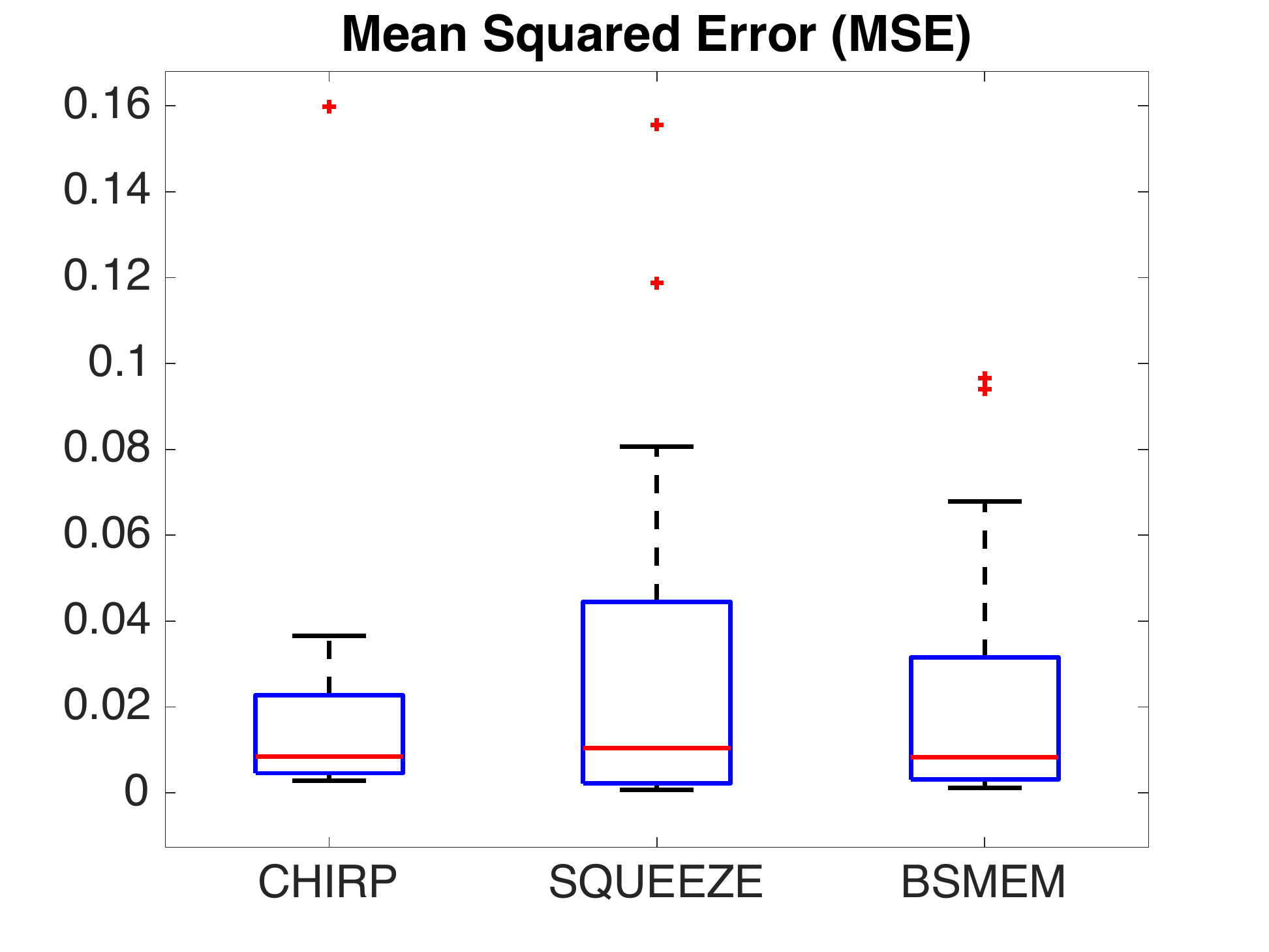}}
	\subfigure{\includegraphics[width=0.49\linewidth]
		{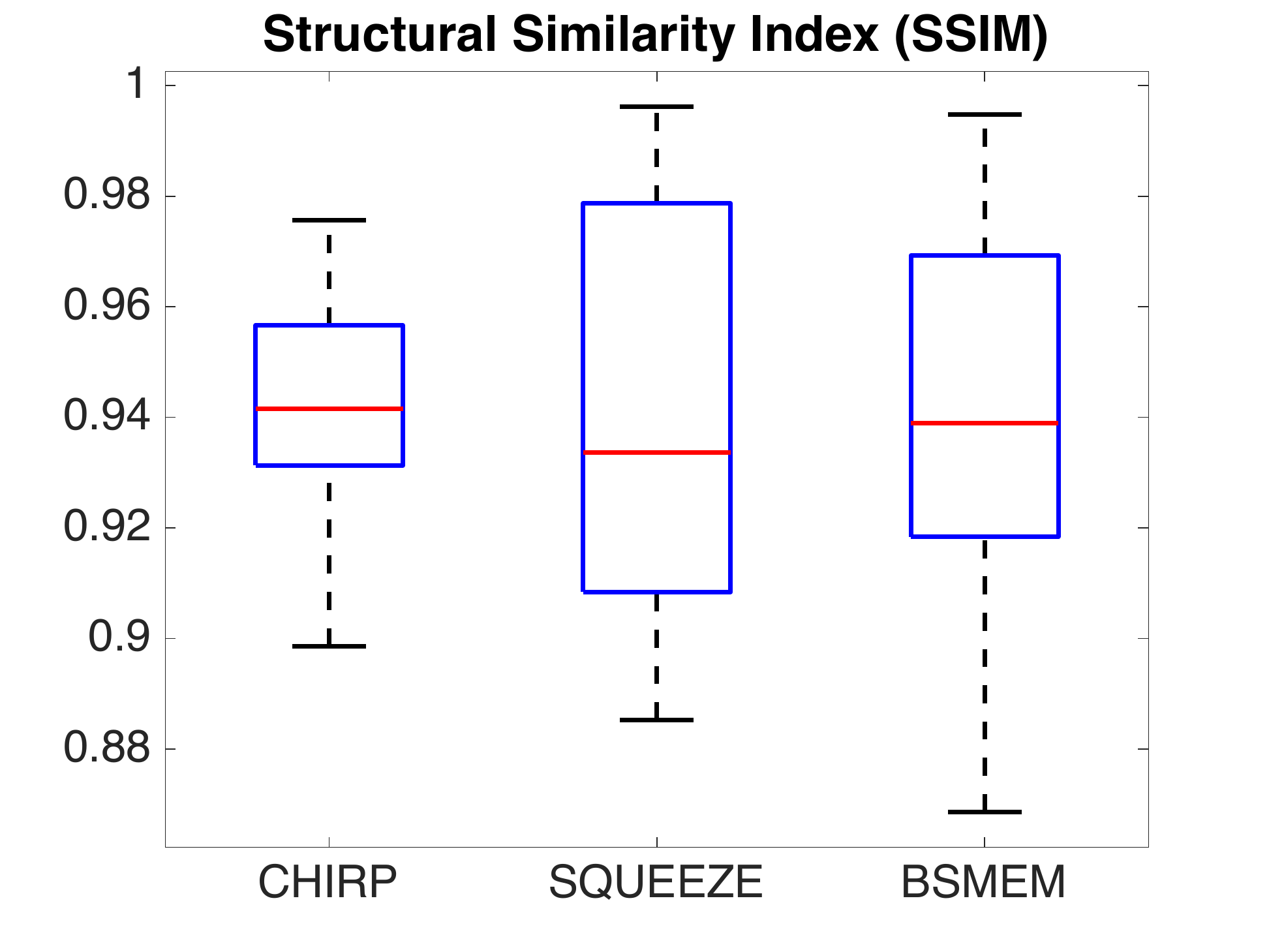}}
	\caption{\footnotesize{ {\bf Quantitative Analysis on Blind Test Set:}  Box plots of MSE and SSIM for reconstruction methods on the blind dataset presented in Section~\ref{section:dataset}. 
			In SSIM a score of 1 implies perceptual indistinguishability between the ground truth and recovered image. Scores are calculated using the original `Source' image (Refer to Fig.~\ref{fig:maxres}). 
		}}
	\label{fig:resultnums}
	\vspace{-0.23in}
\end{figure}

\vspace{-.15in}
\paragraph{Noise Sensitivity} The standard deviation of thermal noise introduced in each measured visibility is fixed based on measurement choices and the corresponding telescopes' properties. Specifically, $\sigma = \frac{1}{0.88} \sqrt{ {\rho_1 \rho_2 }/{ \nu T } } $ for bandwidth $\nu$, integration time $T$, and each telescope's System Equivalent Flux Density ($\rho$).
Consequently, an emission with a lower total flux will result in a lower SNR signal.

Previous measurements predict that the total flux densities of the black holes M87 and SgA* will be in the range 0.5 to 3.0 janskys
~\cite{doeleman2012jet, doeleman2008event}.
Figure~\ref{fig:noise} shows the effect of varying total flux density, and thus noise, on each method's recovered reconstructions. Notice how our method is fairly robust to the noise, while the results from other methods often vary substantially across noise levels.

 \begin{figure}[b]
 	\begin{center}
 		\vspace{-.2in}
 		
%
%
%
%
%
%
%
%

 		\begin{tabular}{c c c c c}

 			\includegraphics[width=.19\linewidth]
 			{blackhole_chirp_1_shift.png} & 	\hspace{-.17in} \includegraphics[width=.19\linewidth]
 			{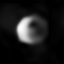} & 	\hspace{-.17in}
 			\includegraphics[width=.19\linewidth]
 			{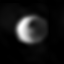}& 	\hspace{-.17in}
 			\includegraphics[width=.19\linewidth]
 			{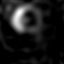}& 	\hspace{-.17in} 
 			\includegraphics[width=.19\linewidth]
 			{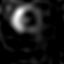} \\

 			 			\includegraphics[width=.19\linewidth]
 			 			{celestial_14_chirp_1.png} & 	\hspace{-.17in} \includegraphics[width=.19\linewidth]
 			 			{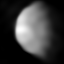} & 	\hspace{-.17in}
 			 			\includegraphics[width=.19\linewidth]
 			 			{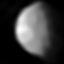}& 	\hspace{-.17in}
 			 			\includegraphics[width=.19\linewidth]
 			 			{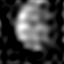}& 	\hspace{-.17in} 
 			 			\includegraphics[width=.19\linewidth]
 			 			{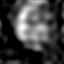} \\
 			
 			\small{\textsf{Natural}}  & \hspace{-.17in} \small{\textsf{Celestial}}  & \hspace{-.17in}
 			\small{\textsf{Black Hole}}  & \hspace{-.17in} \small{\textsf{$\ell_2$ Norm}}  & \hspace{-.17in} \small{\textsf{$\ell_{0.8}$ Norm}} \\
 			

 		\end{tabular}
 		\caption{ \footnotesize{{\bf Effect of Patch Prior:} Reconstructions using patch priors trained on natural, celestial, and synthetic black hole images as well as $\ell_2$ and $\ell_{0.8}$ norm priors on the residuals. The ground truth target image are shown in Figure~\ref{fig:comp} column B and E. 
 				The patch priors outperform results obtained using simpler $\ell$-norm priors. 
 				Since absolute position is lost during imaging, shifts in the reconstructed source location are expected. 
 				} }
 		\label{fig:patchprior}
 		\vspace{-.35in}
 	\end{center}
 \end{figure}

 \vspace{-.15in}
 \paragraph{Effect of Patch Prior} Flexibility of the patch prior framework allows us to easily incorporate a variety of different ``visual" assumptions in our reconstructed image.
 For instance, in the case of the EHT, simulations of a black hole for different inclinations and spins can be used to train a patch model that can be subsequently used for reconstruction. 
 In Figure~\ref{fig:patchprior} we show results using a patch prior trained on natural~\cite{martin2001database}, celestial, and synthetic black hole images
 ~\cite{avery}. Only small variations can be observed among the resulting images.
 Given our selected parameters, this suggests that the prior guides optimization, but does not impose strong assumptions that may greatly bias results.
  

\vspace{-.2in}
\paragraph{Real Measurements} 

We demonstrate the performance of our algorithm on the reconstruction of three different sources using real VLBI data from
~\cite{jorstad2005polarimetric} in Figure~\ref{fig:realdata}.
Although we do not have ground truth images corresponding to these measurements, we compare our reconstructions to those generated by the BU group,
reconstructed using an interactive calibration procedure described in~\cite{jorstad2005polarimetric}.
Alternatively, we are able to use bispectrum measurements to automatically produce image reconstructions with minimal user input. Notice that we are able to recover sharper images, and even resolve two potentially distinct sources that were previously unresolved in blazar OJ287.


\begin{figure}[tb]
	\begin{center}
		\vspace{-.25in}
		\begin{tabular}{  c | c | c | c   }
			&  \small{\textsf{3C279}}  & \small{\textsf{OJ287}}  & \small{\textsf{BL Lacertae}}   \\ \hline
			&\vspace{-.1in}&&\\
			\multirow{1}{*}[0.6in]{ \rotatebox[origin=t]{90}{{\textsf{BU Results}} }}& 
			\includegraphics[width=.2\linewidth]
			{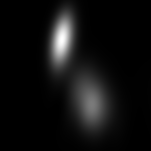} &
			\includegraphics[width=.2\linewidth]
			{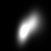} &
			\includegraphics[width=.2\linewidth]
			{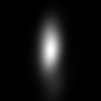}  \\
			
			\hline
			
			&\vspace{-.1in}&&\\
			\multirow{1}{*}[0.5in]{ \rotatebox[origin=t]{90}{{\textsf{CHIRP}} }}& 
			\includegraphics[width=.2\linewidth]
			{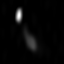} &
			\includegraphics[width=.2\linewidth]
			{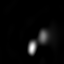} &
			\includegraphics[width=.2\linewidth]
			{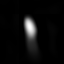}  \\

		\end{tabular}
		\caption{ \footnotesize{{\bf Real Measurements:} A comparison of our reconstructed images to~\cite{jorstad2005polarimetric}'s results using CLEAN self-calibration. Note that we are able to reconstruct less blurry images, and are even able to resolve 2 separate, previously unresolved, bright emissions in blazar OJ287. Measurements were taken using the VLBA telescope array. The FOV for each image is 1.5, 1, and 1 milli-arcsecond respectively.} }
		\label{fig:realdata}
		\vspace{-.3in}
	\end{center}
\end{figure}

\vspace{-.1in}
\section{Conclusion}
\vspace{-.1in}
Astronomical imaging will benefit from the cross-fertilization of ideas with the computer vision community. 
In this paper, we have presented an algorithm, CHIRP, for reconstructing an image using a very sparse number of VLBI frequency constraints.
We have demonstrated improved performance compared to current state-of-the-art methods on both synthetic and real data.
Furthermore, we have introduced a new dataset for the testing and
development of new reconstruction algorithms.  
With this paper, the
dataset, and algorithm comparison website, we hope to make this
problem accessible to researchers in computer vision, and push the limits of celestial imaging. 


\vspace{-0.04in}
\footnotesize{
\paragraph{Acknowledgments}

We would like to thank Andrew Chael, Katherine Rosenfeld, Lindy Blackburn, and Fabien Baron for all of their helpful discussions and feedback. 
This work was partially supported by NSF CGV-1111415.
Katherine Bouman was partially supported by an NSF Graduate Fellowship.
We also thank the National Science Foundation (AST-1310896, AST-1211539, and AST-1440254) and the Gordon and Betty Moore Foundation (GBMF-3561) for financial support of this work.
This study makes use of 43 GHz VLBA data from the VLBA-BU Blazar Monitoring Program (VLBA-BU-BLAZAR;
http://www.bu.edu/blazars/VLBAproject.html), funded by NASA through the Fermi Guest Investigator Program. The VLBA is an instrument of the National Radio Astronomy Observatory. The National Radio Astronomy Observatory is a facility of the National Science Foundation operated by Associated Universities, Inc.
}

	}

\bibliographystyle{ieee}
\bibliography{egbib}

\end{document}